\documentclass[superscriptaddress,nofootinbib,twocolumn,letterpaper,prd]{revtex4-1}

\usepackage{amsmath,amssymb,amsfonts}
\usepackage{mathrsfs}
\usepackage{graphicx}
\usepackage[usenames,dvipsnames]{color}
\usepackage{hyperref}
\usepackage{url}
\usepackage[utf8]{inputenc}
\usepackage{slashed}
\usepackage{subfigure}
\usepackage{slashed,bbm}
\usepackage{graphics,psfrag,epsfig}
\usepackage{dsfont}
\usepackage{setspace}

\usepackage{pdfpages}
\usepackage{pgffor}

\makeatletter
\AtBeginDocument{\let\LS@rot\@undefined}
\makeatother

\graphicspath{{figures/}}

\allowdisplaybreaks

\begin{document}

\title{Emergent symmetries and coexisting orders
in Dirac fermion systems}

\author{Emilio Torres}
\affiliation{Institut f\"ur Theoretische Physik, Universit\"at zu K\"oln, 50937 Cologne, Germany}
\author{Lukas Weber}
\affiliation{Institut f\"ur Theoretische Festk\"orperphysik, JARA-FIT and JARA-HPC,
RWTH Aachen University, 52056 Aachen, Germany}
\author{Lukas Janssen}
\affiliation{Institut f\"ur Theoretische Physik and W\"urzburg-Dresden Cluster of Excellence \textit{ct.qmat}, Technische Universit\"at Dresden, 01062 Dresden, Germany}
\author{Stefan Wessel}
\affiliation{Institut f\"ur Theoretische Festk\"orperphysik, JARA-FIT and JARA-HPC,
RWTH Aachen University, 52056 Aachen, Germany}
\author{Michael~M.~Scherer}
\affiliation{Institut f\"ur Theoretische Physik, Universit\"at zu K\"oln, 50937 Cologne, Germany}

\begin{abstract}
The quantum phase diagram and critical behavior of two-dimensional Dirac fermions coupled to two compatible order-parameter fields with $O(N_1)\oplus O(N_2)$ symmetry is investigated. Recent numerical studies of such systems have reported evidence for non-Landau-Ginzburg-Wilson transitions and emergent \mbox{$O(N_1+N_2)$} symmetry between the two ordered states, which has been interpreted within a scenario of deconfined quantum criticality in (2+1)-dimensional Dirac materials. Here, we provide two theoretical approaches to refine the phase diagrams of such systems. In the immediate vicinity of the multicritical point between the ordered phases and the semimetallic phase, we employ a non-perturbative field-theoretical analysis based on the functional renormalization group. For the particular case of  $N_1=3$, $N_2=1$, we perform a large-scale quantum Monte Carlo analysis of the strong-coupling region, where both orders meet. Our findings support the robust emergence of enhanced symmetry at the multicritical point and suggest the transition between the two ordered phases to take place via a sequence of  continuous transitions. In particular, we find that intermediate regimes of coexistence are present in the phase diagram for all values of $N_1$ and $N_2$.
\end{abstract}

\maketitle

Within  the Landau-Ginzburg-Wilson (LGW) theory of critical phenomena \cite{Landau} a transition between two ordered phases that break different symmetries is either discontinuous or accompanied by a coexistence regime~\cite{Liu1973,PhysRevB.13.412,PhysRevB.8.4270,PhysRevLett.33.813,PhysRevE.78.041124,PhysRevLett.88.059703,PhysRevB.67.054505,Calabrese:2003ia,Kivelson11903,Eichhorn}, unless some fine tuning is performed. A prominent potential exception to this paradigm is the deconfined quantum critical point (DQCP) in spin-$\frac{1}{2}$ antiferromagnets~\cite{Senthil-1}. Within this scenario a quantum critical point separates antiferromagnetic order  from a valence-bond-solid phase, and is  described by  spinon degrees of freedom. These couple to an emergent gauge field and render the transition continuous, while being confined in both of the ordered phases~\cite{Senthil-1,Senthil-2,SenthilJPS,Senthil-VBS}. 
This  DQCP furthermore describes a transition that, according to  numerical evidence~\cite{NahumSerna}, displays an enlarged $O(5)$ symmetry at the critical point. Recent theoretical considerations moreover suggest such emergent $O(N)$ symmetries to be  an ubiquitous feature of deconfined quantum phase transitions~\cite{SernaO4,MengO4,GazitAssaad} and beyond~\cite{RoyO3,ZhangO5,HongYaoSUSY,RoyLorentz,HongYao-FIQCP1,HongYao-FIQCP2,HongYao-FIQCP3,Scherer-FIQCP,KekuleFRGSYM,KekuleFRGSSB}. These ideas may thus be promising also for exploring non-LGW quantum critical fermions.   

Indeed, recent quantum Monte Carlo (QMC) simulations of Dirac fermion systems~\cite{SatoAssaad,Li2019,GazitAssaad} suggest continuous non-LGW transitions between two ordered phases, reminiscent of DQCPs. In particular, the findings in Ref.~\onlinecite{SatoAssaad} for a fermionic model on the  honeycomb lattice indicate that a system of Dirac fermions with anticommuting masses that break an $O(3)$ and $\mathbb{Z}_2$ symmetry, respectively, supports a line of continuous transitions that separates the two phases, featuring an emergent $O(4)$ symmetry. In particular, no definite signs of coexisting orders were reported in Ref.~\onlinecite{SatoAssaad}.

Here, we examine the case of a general system of Dirac fermions coupled to two compatible order parameters (OPs) with $O(N_1)\oplus O(N_2)$ symmetry by following two different and complementary routes: 
a non-perturbative field-theoretical renormalization group (RG) approach, i.e., the functional RG (FRG)~\cite{Wetterich:1992yh}, and a refined QMC analysis for the model in Ref.~\onlinecite{SatoAssaad}. 
The non-perturbative FRG  can be performed directly in $2+1$ dimensions and allows us to assess the multicritical behavior of the model more precisely than leading-order $\epsilon$~expansions~\cite{JHS18,RoyJuricic18,PhysRevB.99.241103}. We firmly establish the emergence of $O(N_1+N_2)$ symmetry at the multicritical point for all consistent values of $N_1$ and $N_2$. Furthermore, our approach facilitates a study of the phases with broken symmetry and we find robust indications for an intermediate coexistence phase for all choices of $N_1,N_2$. Further evidence of coexistence for $N_1=3$, $N_2=1$ is provided by our refined QMC analysis. 

\paragraph*{Effective field theory.}\label{sec:model}

For the FRG analysis, we consider the low-energy effective Gross-Neveu-Yukawa (GNY) model with two OP fields~\cite{PhysRevB.90.041413,PhysRevB.92.035429,Herbut1,Herbut2,PhysRevB.93.125119,JHS18,RoyJuricic18,PhysRevB.99.241103}, describing interacting spin-$1/2$ fermions on the honeycomb lattice in the vicinity of a multicritical point. The Euclidean Lagrangian is $\mathscr{L}=\mathscr{L}_F+\mathscr{L}_B$, where
\begin{align}
\mathscr{L}_F=&\overline{\psi}\left(-i\slashed{\partial}+g_1M_{\phi}+g_2M_{\chi}\right)\psi\,,\nonumber\\
\mathscr{L}_B=&\phi_a\left(-\partial^2+m_1^2\right)\phi_a+\chi_b\left(-\partial^2+m_2^2\right)\chi_b\;\nonumber\\
&+\frac{u_1}{8}\!\left(\phi_a\phi_a\right)^2\!+\!\frac{u_2}{8}\!\left(\chi_b\chi_b\right)^2\!+\!\frac{u_3}{4}\phi_a\phi_a\chi_b\chi_b\;.
\label{eq:lagrangian}
\end{align}
The real OPs $\phi=(\phi_1,...,\phi_{N_1}),\chi=(\chi_1,...,\chi_{N_2})$ act as mass terms $M_{\phi}=\gamma^a_{\phi}\phi_{a}$ and $M_{\chi}=\gamma^b_{\chi}\chi_{b}$, interacting via Yukawa couplings $\bar g_{1/2}$ with eight-component Dirac fermions $\psi$ and $\bar\psi\equiv\psi^\dagger\gamma_0$. Summation over repeated indices is implied. The matrices $\gamma^a_{\phi},\gamma^b_{\chi}$ are defined as $\gamma^a_{\phi}=\gamma_0\beta^a_{\phi}$, 
$\gamma^b_{\chi}=\gamma_0\beta^b_{\chi}$, where the mass matrices $\beta^a_\phi,\beta^b_\chi$ anticommute among each other as well as with the Hamiltonian.
The Lagrangian $\mathscr{L}$ exhibits an effective Lorentz invariance, which is expected to emerge in 2D Dirac fermion systems at criticality~\cite{RoyLorentz, Ray:2018gtp}, and we assume this also in the vicinity of the multicritical point.
The GNY model $\mathscr{L}$  requires $d+N$ anticommuting matrices, where $d$ is the spatial dimension, and $N= N_1+N_2$.
For the eight-dimensional representation relevant for graphene, this implies $N \leq 5$~\cite{RyuShamon, Herbut:2012aa}.

The $O(N_1)\oplus O(N_2)$ symmetric system defined in Eq.~\eqref{eq:lagrangian} includes a symmetry-enlarged $O(N_1+N_2)$-invariant subspace, when $u_1=u_2=u_3$, $m_1^2= m_2^2$, and $g_1^2=g_2^2$.
The leading-order $\epsilon$-expansion analysis~\cite{JHS18,RoyJuricic18} studied the RG fixed point with the enlarged $O(N)$ symmetry,  referred to as the isotropic fixed point (IFP). Within that approach, the IFP is found to be stable for all consistent values of $N$. 
However, it is well known from studies of purely bosonic $O(N_1)\oplus O(N_2)$ models~\cite{PhysRevB.67.054505} that the leading-order $\epsilon$-expansion severely overestimates the stability of the IFP. In contrast, the FRG provides more faithful results already at the low  truncation orders that we exploit here~\cite{Eichhorn}.

An important subtlety in Dirac fermion systems concerns the determination of the nature of a multicritical point. In a purely bosonic theory, a multicritical point can either be bicritical or tetracritical, and the two cases can be distinguished by the sign of the quantity $\Delta=u_1 u_2-u_3^{2}$ in terms of the quartic couplings $u_i$ at the RG fixed point, which is bicritical if $\Delta<0$ and tetracritical if $\Delta>0$~\cite{Liu1973,Eichhorn,PhysRevB.67.054505}. For the symmetry-enhanced case, $\Delta=0$, the above classification is valid if the submanifold in coupling space determined by $\Delta=0$ is closed under the RG flow. This is not generally the case in theories with massless fermions as can be shown using the $\epsilon$-expansion~\cite{JHS18}: the submanifold in theory space defined by $u_1\!=\!u_2\!=\!u_3\!=:\!u\neq0$ satisfies $\Delta=0$, but its dependence on the logarithmic RG scale $t$ is given by $\partial_t\Delta=2 u(g_1^2-g_2^2)^2$.
Therefore, in the presence of Dirac fermions, the sign of $\Delta$ may be subject to change.

\paragraph*{Functional renormalization group.}

We employ the non-perturbative FRG approach~\cite{Wetterich:1992yh,Berges:2000ew} to evaluate the generating functional of one-particle irreducible $n$-point correlation functions $\Gamma^{(n)}$. See Refs.~\cite{Rosa:2000ju,Hofling:2002hj,Gies:2009da,PhysRevD.86.105007,BraunGies11,Scherer:2013pda,Knorr1,Knorr2,Janssen:2014gea,Vacca:2015nta,Feldmann:2017ooy} for applications to low-dimensional GNY systems. This method allows us to calculate directly in $D=2+1$ integrate out the flow also within the ordered phases. 
Central to the FRG method is the exact renormalization group flow equation $\partial_k \Gamma_k = \frac{1}{2}\mathrm{Tr}[\partial_kR_k(\Gamma_k^{(2)}+R_k)^{-1}]$ for the average effective action $\Gamma_k$, where $\Gamma^{(2)}_k$ is the second functional derivative with respect to all field degrees of freedom, and $R_k$ is an infrared cutoff function.
This  flow equation interpolates between the bare action at the ultraviolet (UV) cutoff $k=\Lambda$ and the full quantum effective action $\Gamma=\Gamma_{k=0}$.

We employ a truncation based on the original form of the microscopic action in Eq.~\eqref{eq:lagrangian}, i.e.,
\begin{align}
\Gamma_k&=\int d^Dx \Big\{\overline{\psi} \left(-iZ_{\psi,k}\slashed{\partial}+g_{1,k} M_{\phi}+g_{2,k} M_{\chi}\right)\psi \nonumber\\
&-\frac{1}{2}Z_{\phi,k}\phi_a\partial^2\phi_a-\frac{1}{2}Z_{\chi,k}\chi_b\partial^2\chi_b+V_k(\rho_{\phi},\rho_{\chi})\Big\}\;,
\label{eq:effaction}
\end{align}
which is known as the extended local potential approximation, LPA${}^\prime$. The scale dependence of the Yukawa couplings $g_{1,k}$ and $g_{2,k}$ as well as the wavefunction renormalizations $Z_{\Phi,k}$ have been made explicit.
The bosonic potential $V_k$ depends on the $O(N_1)$ and $O(N_2)$ invariant quantities
$\rho_{\phi}=\phi_{a}\phi_{a}/2$, $\rho_\chi=\chi_{b}\chi_{b}/2$, and we expand its dimensionless form $u=k^{-D}V_k$ as a function of the dimensionless $O(N_i)$-symmetric field variables $\tilde{\rho}_\phi=Z_{\phi,k}k^{2-D}\rho_\phi$ and $\tilde{\rho}_\chi=Z_{\chi,k}k^{2-D}\rho_\chi$ around a (running) minimum, denoted $\kappa_{\phi/\chi}=\tilde\rho_{\phi/\chi,\min}$. A truncation up to order $n_{\max}$ in the fields will be referred to in the following as LPA$^\prime n_{\max}$. The nature of the minimum  corresponds to three different scenarios, depending on whether the individual OP fields are in the symmetric or in the spontaneously symmetry broken regime: (1)~Both OP fields remain in their symmetric phases, i.e., the minimum of the potential is at $\kappa_{\phi/\chi}=0$. (2)~The OP $\phi$ acquires an expectation value, while the other OP $\chi$ remains in its symmetric phase. Similarly, for the  reversed situation. (3)~The minimum of the potential is at nonzero expectation values for both OPs, i.e., $\kappa_{\phi/\chi}\neq0$.

The FRG flow equations for the dimensionless expansion coefficients of the potential $u$, including the quadratic mass terms and the quartic couplings, the dimensionless Yukawa couplings $\tilde{g}^2_{i,k}=g_i^2k^{D-4}Z_{i,k}^{-1}Z_{\psi,k}^{-2}$, as well as the anomalous dimensions $\partial_tZ_{\psi/\phi/\chi}$ are obtained by taking the corresponding derivatives of the ansatz in Eq.~\eqref{eq:effaction}. Flow equations for the expectation values $\kappa_{\phi/\chi}$ are obtained from the condition that they are a minimum, cf. the Supplemental Material~\cite{SM} for details. Essentially, this procedure yields a set of coupled  flow equations of the form $\partial_t x_i=\beta_i(\{x_j\})$ in the space of the  coupling parameters  $x_j$.

\paragraph*{Stability of the symmetry-enhanced fixed point.}

%
\begin{table}[t!]
\begin{tabular*}{\linewidth}{@{\extracolsep{\fill} } c c c c c c c}
\hline\hline
$N_1+N_2$   & $\theta_1$ & $\theta_2$ & $\theta_3$ & $\theta_4$ & $\theta_5$\\ \hline
2 & 0.878 & \textbf{0.864} & \textbf{-0.878} & -1.087 & -1.109\\
3 & \textbf{0.773} & 0.726 & \textbf{-0.924} & -1.179 & -1.322\\
4 & \textbf{0.734} & 0.580 & \textbf{-1.017} & -1.274 & -1.542\\
5 & \textbf{0.738} & 0.465 & \textbf{-1.132} & -1.361 & -1.732\\
\hline\hline
\end{tabular*}
\caption{\label{tab:critexplor} Largest five eigenvalues $\theta_i$ of the stability matrix based on FRG calculation in LPA${}^\prime$8 truncation,  for different values of $N_1+N_2$.  Eigenvalues that already appear in the $O(N_1+N_2)$ symmetric models are printed in boldface.}
\end{table}

A fixed point is defined by the condition $\forall i: \beta_i(\{x_j\})=0$. Here, we focus on the IFP, facilitating the symmetry-enhancement $O(N_1)\oplus O(N_2)\to O(N)$. 
The  system has two tuning parameters represented by the boson masses $m_i^2$, cf. Eq.~\eqref{eq:lagrangian}. They give rise to two relevant directions in the RG flow with scaling properties given by critical exponents, $\theta_1$ and $\theta_2$, obained from the diagonalization of the stability matrix
$\mathcal{M}_{ij}:=-(\partial \beta_i/\partial x_j)\vert_{x^*}$
(here, $x_j^*$  denote the fixed-point coordinates). 
Positive eigenvalues $\theta_i$ of the stability matrix correspond to relevant directions, i.e., their number indicates the number of tuning parameters. Negative $\theta_i$ instead correspond to irrelevant directions, attracted to the fixed point.
The IFP is stable, if the third-largest eigenvalue of the stability matrix is negative (called the \textit{stability exponent}).  

We report the numerical results for the five largest eigenvalues of the stability matrix in Tab.~\ref{tab:critexplor}, exhibiting the stability of the IFP for all consistent choices of $N$. In particular, the stability exponent  has a sizable magnitude of $\mathcal{O}(1)$. Therefore, any perturbation of the enhanced symmetry near the multicritical point will die out rather quickly, supporting a strong tendency towards the emergent symmetry.
This strong tendency towards emergent $O(N_1+N_2)$ is consistent with the  QMC findings in Ref.~\onlinecite{SatoAssaad}. 
Noticeable, the symmetry-enhancement is naturally realized in our extended LGW approach through the fluctuations of massless Dirac fermions, i.e., without requiring the inclusion of additional topological terms.
This contrasts to the case of  purely bosonic $O(N_1)\oplus O(N_2)$ models~\cite{Eichhorn,Calabrese}, for which symmetry-enhancement is  supported only for $N_1=N_2=1$, with an almost marginal  stability exponent.

\begin{figure}[t!]
\begin{center}
\includegraphics[width=\columnwidth]{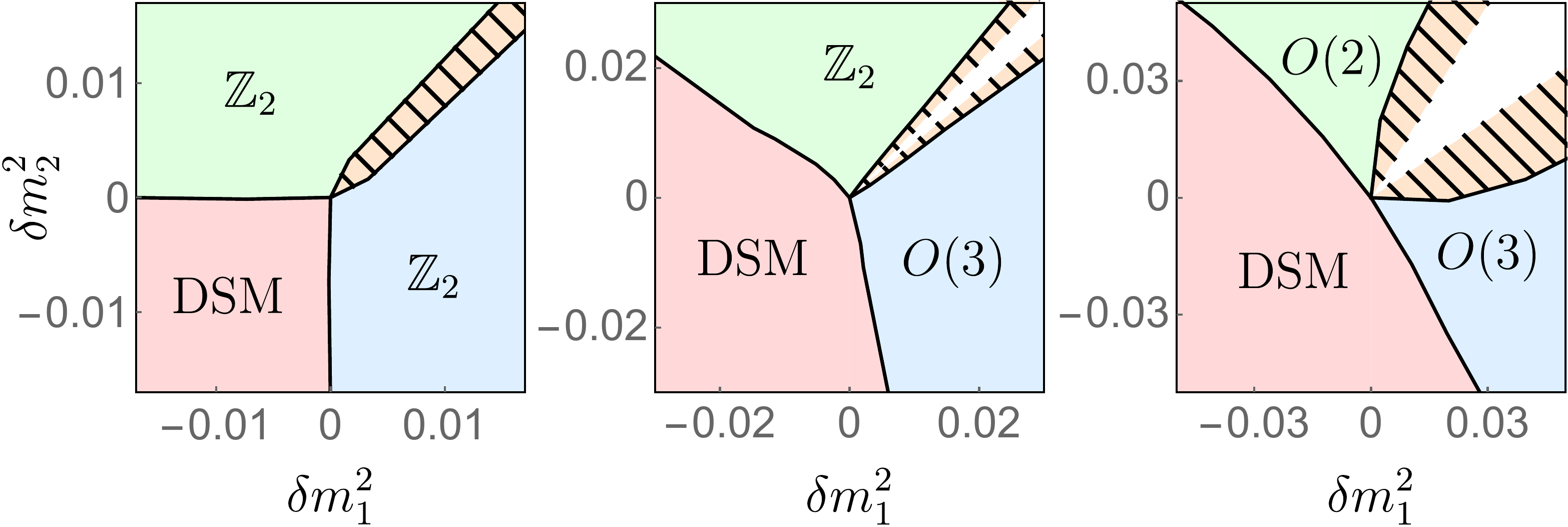}
\end{center}
\caption{Phase diagrams  from FRG for $N_1=1, N_2=1$ (left), $N_1=3, N_2=1$ (middle), and $N_1=3, N_2=2$ (right) near the IFP. Tuning parameters $\delta m_i^2$ measure the distance from the IFP, $\rm{DSM}$ denotes the Dirac semi metal regime, and the other phases are labelled by the broken symmetry.  Solid lines are continuous transitions, and coexistence regions are shown in shaded orange. Within the white areas, the FRG flow is numerically unstable.}
\label{fig:phasediagram}
\end{figure}

\paragraph*{Phase diagram from field theory.}

The phase diagram  finally can be obtained from integrating the FRG flow equations towards the infrared. To this end, we formulate the initial value problem at an arbitrary ultraviolet scale $\Lambda$,  used to set the units. 
To resolve the phase diagram in the vicinity of the symmetry-enhanced quantum multicritical point, perform a sweep of initial conditions in the vicinity if IFP. We consider three different choices for $N_1$ and $N_2$, (i) two coupled Ising order parameters ($N_1 = N_2 = 1$), (ii) a $O(3)\oplus \mathbb{Z}_2$ model ($N_1 = 3$, $N_2 = 1$) -- relevant to the model in Ref.~\onlinecite{SatoAssaad} --  and (iii) a $O(3)\oplus O(2)$ model ($N_1 = 3$, $N_2 = 2$). The resulting phase diagrams are shown in Fig.~\ref{fig:phasediagram}, and we find that all cases exhibit extended coexistence regions, at which both OPs develop a vacuum expectation value.

The lack of data points in the region close to the line $\delta m_1^2=\delta m_2^2$ in Fig.~\ref{fig:phasediagram} emerging from the IFP is related to a numerical instability of the FRG flow. In this region, the expectation values $\kappa_{i,k}$ do not converge to a definite value (either zero or nonzero) in the limit $k\to0$. This behavior occurs along with the appearance of Goldstone modes in each of the adjacent phases, whose interplay with the massive modes has the tendency to drive the system out of the symmetry-broken phase. 
Indeed, when the adjacent symmetry-broken phases involve no massless modes (for $N_1=N_2=1$), the line of exact $O(N)$ symmetry can be resolved effortlessly, while the region of numerical instability grows with the number of available Goldstone modes.
The missing regions of the phase diagrams could be determined by FRG methods  beyond the scope of this paper, e.g., by applying pseudo-spectral methods \cite{Knorr1,Knorr2}. However, the LPA${}^\prime$ truncation already clearly establishes the appearance of a coexistence region near the IFP.

\paragraph*{Quantum Monte Carlo.}
For the case of $O(3)\oplus \mathbb{Z}_2$,
we obtain direct support for the coexistence region also from  a refined QMC analysis for the microscopic  model of Ref.~\onlinecite{SatoAssaad}. For this purpose, we derive an effective quantum spin model that emerges in the strongly-interacting regime of the interacting Dirac fermion model, and which can be simulated by  more efficient QMC methods using cluster updates and  larger lattices than  accessible to the QMC approach for  the original fermionic model. The Hamiltonian of this effective quantum spin model,  obtained by perturbation theory about the strong-interaction limit~\cite{MacDonald88}, reads
\begin{equation}\nonumber
H = \sum_{\langle i,j \rangle} (J_{ij} + \chi_{ij} \sigma_{\langle ij\rangle}^z)\,\mathbf{S}_i \cdot \mathbf{S}_j - J_I \! \!\!\sum_{\langle ij,kl\rangle} \! \!\!\sigma_{\langle ij \rangle}^z  \sigma_{\langle kl\rangle}^z -h \! \sum_{\langle i,j\rangle}\!\sigma_{\langle ij\rangle}^x
\end{equation}
in terms of Heisenberg $S=1/2$ spins $\mathbf{S}_i$, residing on a honeycomb lattice and coupled via bond-centered strengths $\chi_{ij}$ to a transverse-field ($h$) Ising model of  spins  $\mathbf{\sigma}_{\langle ij\rangle}$, located on the nearest-neighbor bonds of the honeycomb lattice. The summation over $\langle i,j\rangle$ ($\langle ij,kl\rangle$) extends over nearest-neighbor Heisenberg (Ising model) spins. In terms of the  nearest-neighbor hopping $t$, the Hubbard interaction $U$, and the fermion-spin couplings $\xi_{ij}$ of the underlying fermionic model, we obtain $J_{ ij} = 4(t^2 + \xi_{ ij}^2)/U$, and $\chi_{ij} = 8 t \xi_{ij}/U$ in second-order perturbation theory~\cite{SM}. These relations  can also be used to specify  parameter values of $H$ and  compare to the results of Ref.~\onlinecite{SatoAssaad}.
\begin{figure}[t!]
\begin{center}
\includegraphics[width=\columnwidth]{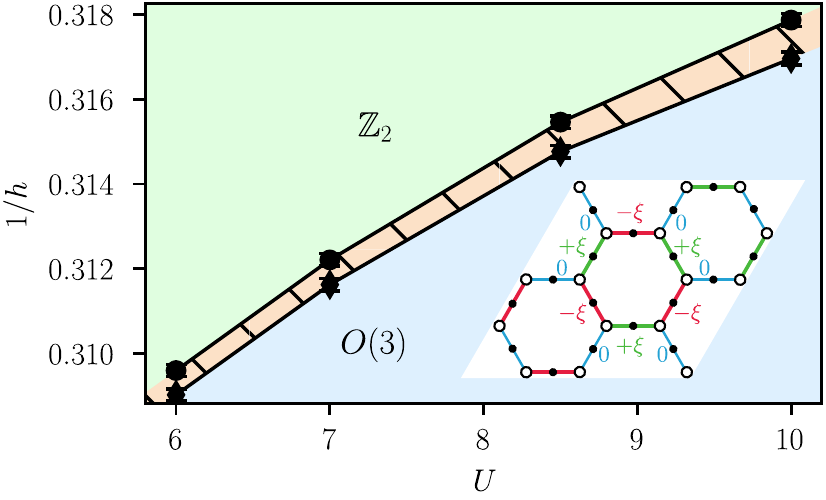}
\end{center}
\caption{QMC phase diagram of the $O(3)\oplus \mathbb{Z}_2$ model $H$ in terms of the parameters $(U,1/h)$ of the underlying fermionic model ($\xi = 0.5$, $J_I = 1$,  $t=1$). The coexistence region is shown in shaded orange. The inset shows the honeycomb lattice with  the modulated pattern $\xi_{ij} = -\xi,0,+\xi$ in different colors. Open (filled) circles denote Heisenberg (Ising) spins. }
\label{fig:qmc1}
\end{figure}

\begin{figure}[t!]
\begin{center}
\includegraphics[width=\columnwidth]{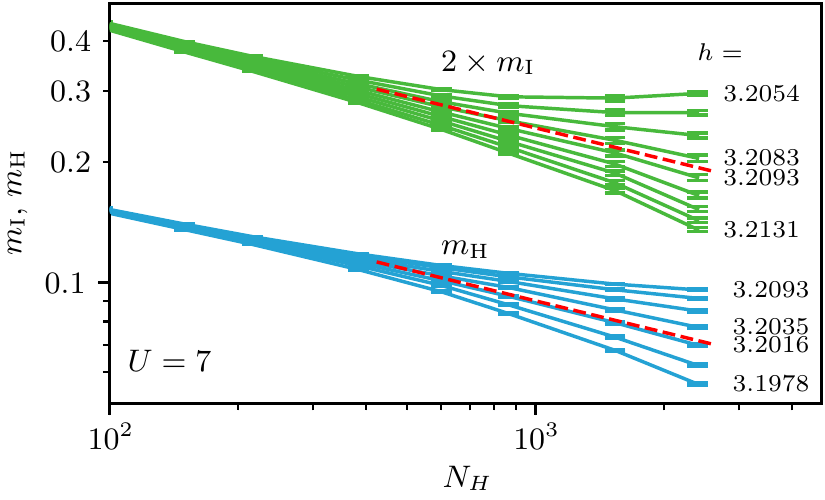}
\end{center}
\caption{Finite-size scaling of $m_\mathrm{H}$ and $m_\mathrm{I}$ for $h$ in the transition region of $H$ for $U=7$ ($\xi = 0.5$, $J_I = 1$,  $t=1$). Dashed  lines show the asymptotic scalings $m_j \propto N_j^{-2\beta_j/\nu_j}$ for $j=\mathrm{I,H}$, with $\beta_j$, $\nu_j$   as of Ref.~\onlinecite{Kos2016}.  For clarity, the values for $m_\mathrm{I}$ were multiplied by a factor of $2$. The lowest and highest shown values of $h$ are given,  as well as the near-critical $h$ values.}
\label{fig:qmc2}
\end{figure}

With a staggered pattern of $\xi_{ij} = +\xi,0,-\xi$ (cf.~the inset of Fig.~\ref{fig:qmc1}) as in Ref.~\onlinecite{SatoAssaad}, the model has a combined lattice-inversion and Ising model spin-flip $\mathbb{Z}_2$ symmetry in addition to the $O(3)$ [$SU(2)$] symmetry of the Heisenberg exchange. To connect  to the previous results, we also fix $\xi = 0.5$, $J_I = 1$, and $t=1$, and  probe the strong-coupling regime for values of $U>6$, where  large single-particle gaps prevailed. We used a hybrid QMC parallel tempering scheme~\cite{SM,PT1,PT2} for the Hamiltonian $H$ on  periodic lattices with $N_\mathrm{H}$ Heisenberg spins (and $N_\mathrm{I}=3N_\mathrm{H}/2$ Ising spins), for $N_\mathrm{H}$ up to 2400, based on the stochastic series expansion approach~\cite{Sandvik99,Sandvik03,Sengupta02}. 
 In particular, we monitored the evolution of the ferromagnetic Ising OP 
 $m_\mathrm{I}=\langle | \frac{1}{N_\mathrm{I}}\sum_{b=1}^{N_\mathrm{I}} \sigma^z_b | \rangle$ 
 and the antiferromagnetic Heisenberg OP
 $m_\mathrm{H}=\langle | \frac{1}{N_\mathrm{H}}\sum_{i=1}^{N_\mathrm{H}}  (-1)^ i S^z_i | \rangle$ 
 upon varying $h$. 
 As an example, Fig.~\ref{fig:qmc2}  shows the finite-size scaling  of both OPs, for $h$ in the transition region at $U=7$. The algebraic behaviours at the order-to-disorder transitions  are in accord with the  anticipated  asymptotic scalings for the purely bosonic Ising and Heisenberg universality classes in dimension $D=2+1$, and yield two distinct  critical field strengths, 
$h_c^\mathrm{I}=3.2088(5)$ for $m_\mathrm{I}$, and $h_c^\mathrm{H}=3.202(2)$ for $m_\mathrm{H}$. 
 
The resulting phase diagram in the above parameter regime is shown in Fig.~\ref{fig:qmc1}. It exhibits a gosshamer intermediate phase of coexisting orders between the small (large) $h$ phase with pure $ \mathbb{Z}_2$ ($O(3)$) symmetry breaking:
For small $h$, the Ising model spins order ferromagnetically, spontaneously breaking the $\mathbb{Z}_2$ symmetry. Due to the staggered $\xi_{ ij}$-coupling, this induces a preferred dimerization pattern on the honeycomb lattice, which leads to a dominant valence-bond singlet formation of the Heisenberg spins along the stronger ($\xi_{ij} \sigma_{\langle ij \rangle}^z>0$) bonds. Upon increasing $h$, the ferromagnetic  order reduces, in effect weakening also   
 the induced dimerization, so that antiferromagnetic Heisenberg spin order can eventually set in prior to the full suppression of the ferromagnetic Ising spin order at even larger values of $h$.
Due to the large single-particle gaps for $U>6$, we  exclude   residual (charge) fluctuations in the fermionic 
model to qualitatively modify this basic physics. We observe in Fig.~\ref{fig:qmc1} that the coexistence regime widens little with increasing  $U$.
This  thinness of the coexistence regime within the considered parameter regime explains, why it was not resolved by the fermionic QMC methods~\cite{SatoAssaad}.   It should however be noted, that within the effective quantum spin model we cannot  identify the multicritical point of the underlying fermionic theory, which was located at $(U,1/h)\approx (4.2, 0.28)$ in  Ref.~\onlinecite{SatoAssaad}. Indeed, the multicritical IFP  is of genuine fermionic nature, cf. the FRG results. 

\paragraph*{Discussion.}

We studied $(2+1)$-dimensional Dirac fermions coupled to two compatible OPs with $O(N_1)$ and $O(N_2)$ symmetry in the vicinity of the multicritical isotropic fixed point, providing an emergent $O(N_1+N_2)$ symmetry. Our FRG study predicts a strong irrelevance of perturbations of the $O(N_1+N_2)$ symmetry, which is consistent with the numerical result of an emergent symmetry in the related Dirac fermion lattice model~\cite{SatoAssaad}. 
Contrary to this  numerical study, however, we do not find a single line of direct, continuous order-to-order transitions. Instead, we identify a robust region of coexistence of both orders, which is separated by continuous transitions from the other phases. In the case of the  $O(3)\oplus\mathbb{Z}_2$ symmetry, relevant for Ref.~\onlinecite{SatoAssaad}, we could furthermore support this field-theoretical analysis by large-scale QMC simulations. Such a combined approach should  be fruitful for uncovering the nature of quantum critical fermions coupled to bosonic fields under a wide range of conditions also in related systems. 

\paragraph*{Acknowledgments.}
The authors are grateful to F.~F.~Assaad and I.~F.~Herbut for discussions. 
E.T. was supported by the Deutsche Forschungsgemeinschaft  (DFG) through the Leibniz Prize of A. Rosch, RO2265/5-1.
L.J. was supported by the DFG through JA2306/4-1 (Emmy Noether program, project id 411750675), SFB 1143 (project id 247310070), and the W\"urzburg-Dresden Cluster of Excellence \textit{ct.qmat} (EXC 2147, project id 39085490). M.M.S. was supported by the DFG, Projektnummer 277146847 -- SFB 1238 (project C02, C03).
L.W. and S.W. acknowledge support by DFG under Grant No. WE/3649/4-2 of the FOR 1807 and through RTG 1995. Furthermore, we acknowledge the IT Center at RWTH Aachen University and the JSC J\"ulich for access to computing time through JARA-HPC.

\bibliography{CompOps}



\section*{Supplementary material}

# Supplemental Material:
# Emergent symmetries and coexisting orders in Dirac fermion systems

Emilio Torres,[1] Lukas Weber,[2] Lukas Janssen,[3] Stefan Wessel,[2] and Michael M. Scherer[1]

[1]*Institut für Theoretische Physik, Universität zu Köln, 50937 Cologne, Germany*
[2]*Institut für Theoretische Festkörperphysik, JARA-FIT and JARA-HPC,
RWTH Aachen University, 52056 Aachen, Germany*
[3]*Institut für Theoretische Physik and Würzburg-Dresden Cluster of Excellence ct.qmat,
Technische Universität Dresden, 01062 Dresden, Germany*

## I. FUNCTIONAL RENORMALIZATION GROUP FLOW EQUATIONS

The central object of the FRG method is the exact renormalization group flow equation for the average effective action $\Gamma_k$, reading [1]

$$\partial_t \Gamma_k = \frac{1}{2}\mathrm{STr}[\partial_t R_k (\Gamma_k^{(2)} + R_k)^{-1}] \,, \tag{1}$$

where $\Gamma_k^{(2)}$ is the Hessian in field space and $R_k$ is an infrared cutoff function, described in detail below. We employ the notation $t = \ln(k/\Lambda)$ for the renormalization group time. The graded trace STr involves a sum over continuous degrees of freedom as well as spinor indices, the latter entering with a minus sign. This exact evolution equation can be approximated by employing an ansatz based on the original form of the microscopic action, i.e.,

$$\Gamma_k = \int d^D x \left\{ \overline{\psi}_\nu \left(-i Z_{\psi,k} \slashed{\partial} + g_{1,k} M_\phi + g_{2,k} M_\chi \right) \psi_\nu - \frac{1}{2} Z_{\phi,k} \phi_a \partial^2 \phi_a - \frac{1}{2} Z_{\chi,k} \chi_b \partial^2 \chi_b + V_k(\phi_a, \chi_b) \right\} \,. \tag{2}$$

where we use the summation convention on repeated indices. This truncation is known as the extended local potential approximation or LPA′. Furthermore we assume that the bosonic potential $V_k(\phi,\chi)$ is an analytic function of the $O(N_1)$ and $O(N_2)$ invariant quantities

$$\rho_\phi := \phi_a \phi_a / 2, \qquad \rho_\chi := \chi_b \chi_b / 2 \,. \tag{3}$$

Anticipating the splitting of the bosonic degrees of freedom into longitudinal and transversal, we denote the combined independent bosonic fields as $\Phi := \{\phi_L, \phi_T, \chi_L, \chi_T\}$. Moreover, we make use of the optimized linear regulator functions for both, bosons and fermions [2–4], i.e.,

$$R_{\Phi,k}(p) = Z_{\Phi,k} p^2 r_B(p^2/k^2)\,, \tag{4}$$
$$R_{\psi,k}(p) = -Z_{\psi,k} \slashed{p}\, r_F(p^2/k^2)\,, \tag{5}$$

where

$$r_B(x) = \left(\frac{1}{x} - 1\right) \Theta(1-x)\,, \tag{6}$$
$$r_F(x) = \left(\frac{1}{\sqrt{x}} - 1\right) \Theta(1-x)\,, \tag{7}$$

and $\Theta(x)$ is a step function. With these choices, we then feed the Ansatz of Eq. (2) into the Wetterich equation Eq. (1). The respective projections are described below, where to alleviate the notation, and unlike in the main text, we denote dimensionful quantities with a tilde.

### A. Masses and bosonic potential

The different masses appearing in the threshold equations are obtained from the Hessian of the dimensionless potential $u(\rho_\phi, \rho_\chi)$. Denoting derivatives with respect to $\Phi$ with subscripts and derivatives with respect to $\rho_\phi, \rho_\chi$ with superscripts, (i.e., $u^{(m,n)} := \frac{\partial^{m+n} u}{\partial \rho_\phi^m \partial \rho_\chi^n}$), the nonzero entries are given by

$$u_{11} = u^{(1,0)} + 2\rho_\phi u^{(2,0)}\,,\quad u_{22} = u^{(1,0)}\,,\quad u_{33} = u^{(0,1)} + 2\rho_\chi u^{(0,2)}\,,\quad u_{44} = u^{(0,1)}\,,\quad u_{13} = u_{31} = 2\sqrt{\rho_\phi \rho_\chi}\, u^{(1,1)}\,,$$

and $\omega_\psi = 2(g_1^2 \rho_\phi + g_2^2 \rho_\chi)$. From this we define the entries of the bosonic propagator matrix as

$$d_{1/2,L} := \frac{1 + u_{33/11}}{(1+u_{11})(1+u_{33}) - u_{13}^2}, \quad d_{iT} := \frac{1}{1+u_{2i2i}}, \quad d_{\phi\chi} := \frac{u_{13}}{u_{13}^2 - (1+u_{11})(1+u_{33})}. \tag{8}$$

The flow equation for the potential takes the form

$$\partial_t u = -Du + (D - 2 + \eta_\phi)\rho_\phi u^{(1,0)} + (D - 2 + \eta_\chi)\rho_\chi u^{(0,1)} - d_\gamma \ell^F + \ell^B_{1L} + (N_1 - 1)\ell^B_{1T} + \ell^B_{2L} + (N_2 - 1)\ell^B_{2T} \tag{9}$$

with threshold functions

$$\ell^B_{i\alpha} := \frac{4v_D}{D}\left(1 - \frac{\eta_i}{D+2}\right) d_{i\alpha}, \quad \ell^F := \frac{4v_D}{D}\left(1 - \frac{\eta_\psi}{D+1}\right)\frac{1}{1+\omega_\psi}. \tag{10}$$

The potential $u$ can be expanded in different ways corresponding to three different scenarios depending on whether the order-parameter fields are in the symmetric (SYM) or in the spontaneously symmetry broken regime (SSB):

(1) SYM-SYM: Both OP fields remain in their symmetric phases, i.e., the minimum of the potential is at $\kappa_{\phi/\chi} = 0$. In this case, the potential can be expanded as

$$u(\rho_\phi, \rho_\chi) = m_\phi^2 \rho_\phi + m_\chi^2 \rho_\chi + \sum_{m+n=2}^{n_{\max}/2} \frac{\lambda_{mn}}{m!n!} \rho_\phi^m \rho_\chi^n. \tag{11}$$

(2) SSB-SYM: The OP $\phi$ acquires an expectation value, while the other OP $\chi$ remains in its symmetric phase.

$$u(\rho_\phi, \rho_\chi) = m_\chi^2 \rho_\chi + \sum_{m+n=2}^{n_{\max}/2} \frac{\lambda_{mn}}{m!n!}(\rho_\phi - \kappa_\phi)^m \rho_\chi^n. \tag{12}$$

The related situation with the reversed role of the two OPs is referred to as SYM-SSB.

(3) SSB-SSB: The minimum of the potential is at nonzero expectation values for both OPs, i.e., $\kappa_{\phi/\chi} \neq 0$.

$$u(\rho_\phi, \rho_\chi) = \sum_{m+n=2}^{n_{\max}/2} \frac{\lambda_{mn}}{m!n!}(\rho_\phi - \kappa_\phi)^m (\rho_\chi - \kappa_\chi)^n. \tag{13}$$

### B. Projections on couplings

Equations for the bosonic couplings are obtained from Eq. (9) by taking the appropriate derivatives. In the SYM-SYM phase, we have

$$\partial_t \lambda_{m,n} = \partial_t u^{(m,n)}\Big|_{\rho_\phi = \rho_\chi = 0}. \tag{14}$$

In the SSB-SYM case, we have

$$\partial_t \lambda_{m,n} = \left(\partial_t u^{(m,n)} + u^{(m+1,n)}\partial_t \kappa_\phi\right)\Big|_{\rho_\phi = \kappa_\phi, \rho_\chi = 0},$$

$$\partial_t \kappa_\phi = -\frac{\partial_t u^{(1,0)}}{u^{(2,0)}}\Big|_{\rho_\phi = \kappa_\phi, \rho_\chi = 0}, \tag{15}$$

and finally, in the SSB-SSB case, we need to calculate

$$\partial_t \lambda_{m,n} = \left(\partial_t u^{(m,n)} + u^{(m+1,n)}\partial_t \kappa_\phi + u^{(m,n+1)}\partial_t \kappa_\chi\right)\Big|_{\rho_\phi = \kappa_\phi, \rho_\chi = \kappa_\chi}. \tag{16}$$

In the latter case, the location of the minimum of the potential flows according to

$$\partial_t \kappa_\phi = \frac{u^{(0,2)}\partial_t u^{(1,0)} - u^{(1,1)}\partial_t u^{(0,1)}}{(u^{(1,1)})^2 - u^{(2,0)}u^{(2,0)}}\Big|_{\rho_\phi = \kappa_\phi, \rho_\chi = \kappa_\chi},$$

$$\partial_t \kappa_\chi = \frac{u^{(2,0)}\partial_t u^{(0,1)} - u^{(1,1)}\partial_t u^{(1,0)}}{(u^{(1,1)})^2 - u^{(2,0)}u^{(2,0)}}\Big|_{\rho_\phi = \kappa_\phi, \rho_\chi = \kappa_\chi}.$$

### C. Yukawa couplings

To be able to compare with results from the perturbative $\epsilon$-expansion results, we take a projection along the longitudinal directions of the OPs. The coordinates are chosen such that $\phi_1$ and $\chi_1$ are the massive directions, this means

$$\partial_t \tilde{g}_i = \frac{1}{d_\gamma} \text{tr} \left( \gamma_\Phi^1 \frac{\delta}{\delta \Phi_{2i-1}(q)} \frac{\overrightarrow{\delta}}{\delta \bar{\psi}(q)} \partial_t \Gamma_k \frac{\overleftarrow{\delta}}{\delta \psi(q)} \right) \Bigg|_{\substack{q=0, \Phi=0 \\ \bar{\psi}=\psi=0}}.$$

The full flow equations thus obtained are

$$\partial_t g_1^2 = (D - 4 + \eta_1 + 2\eta_\psi) g_1^2 - 2g_1^2 \Big[ g_1^2 \Big( (N_1 - 1) L_{110,1T}^{FB} - L_{111,1L}^{FB} \Big) + g_2^2 \Big( (N_2 - 1) L_{110,2T}^{FB} + L_{111,2L}^{FB} \Big) \Big] \quad (18)$$
$$+ 2g_1^2 \Big[ u_{221}(N_1 - 1) \sqrt{2\rho_\phi} g_1^2 L_{120,1T}^{FB} + u_{441}(N_2 - 1) \sqrt{2\rho_\chi} g_2^2 L_{120,2T}^{FB} \Big]$$
$$- 2g_1^2 u_{111} \Big[ \sqrt{2\rho_\phi} \Big( g_1^2 L_{122,1L}^{FB} - g_2^2 L_{12}^{BB} \Big) + 2g_2^2 \sqrt{2\rho_\chi} L_{12,1}^{FR} \Big] - 2g_1^2 u_{331} \Big[ \sqrt{2\rho_\phi} \Big( g_1^2 L_{12}^{BB} - g_2^2 L_{122,2L}^{FB} \Big) + 2g_2^2 \sqrt{2\rho_\chi} L_{12,2}^{FR} \Big]$$
$$- 2g_1^2 (u_{131} + u_{311}) \Big[ \sqrt{2\rho_\phi} \Big( g_1^2 L_{12,1}^{FR} - g_2^2 L_{12,2}^{FR} \Big) + g_2^2 \sqrt{2\rho_\chi} (L_{11}^{BR} + L_{02}^{BR}) \Big] - 16 g_1^4 g_2^2 \sqrt{\rho_\phi \rho_\chi} L_{21}^{BB}$$
$$- 8g_1^4 \rho_\phi \Big[ g_1^2 (L_{211,1L}^{FB} - (N_1 - 1) L_{210,1T}^{FB}) - g_2^2 ((N_2 - 1) L_{210,2T}^{FB} + L_{211,2L}^{FB}) \Big],$$

and

$$\partial_t g_2^2 = (D - 4 + \eta_2 + 2\eta_\psi) g_2^2 - 2g_2^2 \Big[ g_2^2 \Big( (N_2 - 1) L_{110,2T}^{FB} - L_{111,2L}^{FB} \Big) + g_1^2 \Big( (N_1 - 1) L_{110,1T}^{FB} + L_{111,1L}^{FB} \Big) \Big] \quad (19)$$
$$+ 2g_2^2 \Big[ u_{443}(N_2 - 1) \sqrt{2\rho_\chi} g_2^2 L_{120,2T}^{FB} + u_{223}(N_1 - 1) \sqrt{2\rho_\phi} g_1^2 L_{120,1T}^{FB} \Big]$$
$$- 2g_2^2 u_{333} \Big[ \sqrt{2\rho_\chi} \Big( g_2^2 L_{122,2L}^{FB} - g_1^2 L_{12}^{BB} \Big) + 2g_1^2 \sqrt{2\rho_\phi} L_{12,2}^{FR} \Big] - 2g_2^2 u_{113} \Big[ \sqrt{2\rho_\chi} \Big( g_2^2 L_{12}^{BB} - g_1^2 L_{122,1L}^{FB} \Big) + 2g_1^2 \sqrt{2\rho_\phi} L_{12,1}^{FR} \Big]$$
$$- 2g_2^2 (u_{313} + u_{133}) \Big[ \sqrt{2\rho_\chi} \Big( g_2^2 L_{12,2}^{FR} - g_1^2 L_{12,1}^{FR} \Big) + g_1^2 \sqrt{2\rho_\phi} (L_{11}^{BR} + L_{02}^{BR}) \Big] - 16 g_2^4 g_1^2 \sqrt{\rho_\phi \rho_\chi} L_{21}^{BB}$$
$$- 8g_2^4 \rho_\chi \Big[ g_2^2 (L_{211,2L}^{FB} - (N_2 - 1) L_{210,2T}^{FB}) - g_1^2 ((N_1 - 1) L_{210,1T}^{FB} + L_{211,1L}^{FB}) \Big],$$

with threshold functions given by

$$L_{mnr,i\alpha}^{FB} := \frac{8 v_D}{D} \left[ \left( 1 - \frac{\eta_\psi}{D+1} \right) \frac{m d_{i\alpha}}{1 + \omega_\psi} + \left( 1 - \frac{\eta_i}{D+2} \right) n d_{i\alpha}^2 + \left( 1 - \frac{\eta_j}{D+2} \right) r |\epsilon_{ij}| d_{\phi\chi}^2 \right] \frac{d_{i\alpha}^{n-1}}{(1 + \omega_\psi)^m}, \quad (20)$$

$$L_{mn,i}^{FR} := \frac{8 v_D}{D} \left[ \left( 1 - \frac{\eta_\psi}{D+1} \right) \frac{m d_{iL}}{1 + \omega_\psi} + \left( 1 - \frac{\eta_i}{D+2} \right) n d_{iL}^2 + \left( 1 - \frac{\eta_j}{D+2} \right) |\epsilon_{ij}| (d_{\phi\chi}^2 + d_{1L} d_{2L}) \right] \frac{d_{\phi\chi}}{(1 + \omega_\psi)^m},$$

$$L_{mn}^{BB} := \frac{8 v_D}{D} \left[ \left( 1 - \frac{\eta_\psi}{D+1} \right) \frac{m}{1 + \omega_\psi} + n \left( \left( 1 - \frac{\eta_1}{D+2} \right) d_{1L} + \left( 1 - \frac{\eta_2}{D+2} \right) d_{2L} \right) \right] \frac{d_{\phi\chi}^n}{(1 + \omega_\psi)^m},$$

$$L_{mn}^{BR} := \frac{8 v_D}{D} \left[ \left( 1 - \frac{\eta_\psi}{D+1} \right) \frac{m}{1 + \omega_\psi} + n \left( \left( 1 - \frac{\eta_1}{D+2} \right) d_{1L} + \left( 1 - \frac{\eta_2}{D+2} \right) d_{2L} \right) \right] \frac{d_{\phi\chi}^2 + m d_{1L} d_{2L}}{1 + \omega_\psi},$$

where the index $i, j \in \{1, 2\}$ referring to the two distinct Yukawa couplings is summed over when repeated and $\alpha \in \{L, T\}$ refers to the longitudinal or transverse components, respectively.

### D. Anomalous dimensions

Finally, the set of equations is closed by considering the anomalous dimensions. A projection along the longitudinal components leads to

$$\partial_t Z_i = \lim_{q \to 0} \frac{\partial}{\partial q^2} \frac{\overrightarrow{\delta}}{\delta \Phi_{2i}(q)} \partial_t \Gamma_k \frac{\overleftarrow{\delta}}{\delta \Phi_{2i}(-q)} \Bigg|_{\Phi = \bar{\psi} = \psi = 0},$$

$$\partial_t Z_\psi = -\lim_{q\to 0} \frac{1}{d_\gamma D} \operatorname{tr}\left(\gamma^\mu \frac{\partial}{\partial q^\mu} \frac{\overrightarrow{\delta}}{\delta\bar\psi(q)} \partial_t \Gamma_k \frac{\overleftarrow{\delta}}{\delta\psi(q)}\right)\bigg|_{\substack{\Phi=0 \\ \bar\psi=\psi=0}},$$

from which one gets

$$\eta_i = \frac{4v_D}{D}\left[m_i^B + 2d_\gamma g_i^2 m_i^F\right], \tag{21}$$

$$\eta_\psi = \frac{8v_D}{D}\left(g_1^2\left(m_{11,1L}^{FB} + (N_1-1)m_{10,1T}^{FB}\right) + g_2^2\left(m_{11,2L}^{FB} + (N_2-1)m_{10,2T}^{FB}\right)\right), \tag{22}$$

with

$$m_i^F := \left(\frac{1-\eta_\psi}{D-2}+1\right)\frac{2}{(1+\omega_\psi)^3} - \left(\frac{1-\eta_\psi}{D-2}+\frac{1}{2}\right)\frac{1}{(1+\omega_\psi)^2} - \frac{4g_i^2\rho_i}{(1+\omega_\psi)^4}, \tag{23}$$

$$m_1^B := \left((d_{1L}^2 + d_{\phi\chi}^2)u_{111} + (d_{1L}+d_{2L})d_{\phi\chi}u_{131}\right)^2 + \left((d_{2L}^2 + d_{\phi\chi}^2)u_{331} + (d_{1L}+d_{2L})d_{\phi\chi}u_{311}\right)^2$$
$$+ 2\left((d_{1L}^2 + d_{\phi\chi})^2)u_{131} + (d_{1L}+d_{2L})d_{\phi\chi}u_{111}\right)\left((d_{2L}^2 + d_{\phi\chi})^2)u_{131} + (d_{1L}+d_{2L})d_{\phi\chi}u_{331}\right)$$
$$+ (N_1-1)u_{221}^2 d_{1T}^4 + (N_2-1)u_{441}^2 d_{2T}^4, \tag{24}$$

$$m_2^B := \left((d_{2L}^2 + d_{\phi\chi}^2)u_{333} + (d_{1L}+d_{2L})d_{\phi\chi}u_{313}\right)^2 + \left((d_{1L}^2 + d_{\phi\chi}^2)u_{113} + (d_{1L}+d_{2L})d_{\phi\chi}u_{133}\right)^2$$
$$+ 2\left((d_{2L}^2 + d_{\phi\chi})^2)u_{313} + (d_{1L}+d_{2L})d_{\phi\chi}u_{333}\right)\left((d_{1L}^2 + d_{\phi\chi})^2)u_{313} + (d_{1L}+d_{2L})d_{\phi\chi}u_{113}\right)$$
$$+ (N_1-1)u_{223}^2 d_{1T}^4 + (N_2-1)u_{443}^2 d_{2T}^4, \tag{25}$$

$$m_{mn,i\alpha}^{FB} := \left(m\left(1-\frac{\eta_i}{D+1}\right)d_{i\alpha}^2 + n\left(1-\frac{\eta_j}{D+1}\right)|\epsilon_{ij}|d_{\phi\chi}^2\right)\frac{1}{1+\omega_\psi}. \tag{26}$$

## II. PHASES FROM THE FIELD THEORY

The RG flow diagram in the relevant subsector, corresponding to the bosonic masses $m_1^2$ and $m_2^2$ with all other couplings set to their fixed point values, can be seen in Fig. 1 for $N = 5$. Arrows indicate the flow towards the infrared

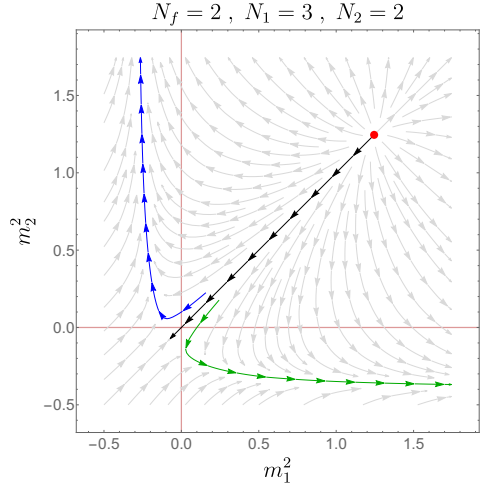

FIG. 1. **Flow diagram in the RG relevant submanifold** spanned by $(m_1^2, m_2^2)$ for $(N_1, N_2) = (3,2)$ when all other couplings are set to their IFP values within the LPA4$'$. The red dot corresponds to the IFP, and the black stream line to a flow within the enlarged-symmetry subspace. Stream lines that cross the vertical (horizontal) axis (red lines) correspond to flows where the $O(N_1)$ ($O(N_2)$) symmetry breaks.

5and the situation is qualitatively the same for all other combinations of $N_1$ and $N_2$ with $N = N_1 + N_2 \leq 5$. Negative values of $m_i^2$, $i \in \{1, 2\}$ correspond to a phase where the $O(N_i)$ symmetry is spontaneously broken and thus the flow diagram Fig. 1 suggests a tendency towards phases where a single one of the symmetries spontaneously broken and the other one remains unbroken. This is represented by flows where $m_1^2 < 0$ stay in the $m_2^2 > 0$ region or vice versa, as indicated by the blue (green) stream lines. This preliminary analysis is incomplete for several reasons. First, the flows for $m_{i,k}^2 < 0$ need to take into account that the minimum of the bosonic potential lies away from the origin in field space, a fact which is neglected in Eq. (11). Additionally, these streams can be misleading in the sense that there appears to be a saddle point for $m_1^2 = m_2^2 < 0$. Upon closer examination this is, in fact, not a fixed point of the whole set of flow equations.

To expand on this preliminary analysis, one needs to integrate the full FRG flows. We show a paradigmatic FRG flow in Fig. 2. In the upper panel, where the flow evolves through several regimes, as described above, and we switch the parametrization of the bosonic potential accordingly. We note that there is still a residual RG running in the deep infrared due to the presence of Goldstone modes in the SSB regimes, see also, for example, Refs. 5–7. For comparison, we also show the flow of the quartic couplings in the lower panel of Fig. 2.

Identifying the low-energy phase of the system for a particular choice of the initial conditions requires determining the regime of the effective potential which is adopted in the deep infrared, i.e., for $k \to 0$, or equivalently, $t \to -\infty$. To that end, we follow the values of the expectation values of the OPs, e.g.,

$$\lim_{k \to 0} \kappa_\phi \quad \text{and} \quad \lim_{k \to 0} \kappa_\chi \,. \tag{27}$$

In Fig. 2 both expectation values $\kappa_i$ are finite in the deep infrared and the flow ends up in the coexistence phase, where both $O(N_i)$ symmetries are spontaneously broken. Similar flow trajectories like the one presented in Fig. 2 are found for other initial conditions and by systematically scanning parameter space in the vicinity of the IFP, we confirm that a system described by the effective theory of Eq. (2) indeed realizes all three possibilities mentioned around Eqs. (11)–(13).

The phase where $\lim_{k\to 0} \kappa_{\phi,k} = \lim_{k\to 0} \kappa_{\chi,k} = 0$, i.e., where both symmetries are left unbroken, is connected to the Dirac semimetallic (DSM) phase. This phase can be further characterized in LPA∞′, i.e., to any order in the LPA′, by nonvanishing universal fixed point values for all running couplings [6]. This means, in particular, that the Yukawa and quartic couplings satisfy $g_1^2 = g_2^2 =: g^2/v_D$ and $\lambda_{2,0} = \lambda_{1,1} = \lambda_{0,2} =: \lambda/v_D$, and they flow to the universal values

$$g_{\text{DSM},*}^2 = \frac{D(4-D)(D-2)}{16(3D-4)} \,, \tag{28}$$

$$\lambda_{\text{DSM},*} = \frac{D(4-D)(D-2)^2}{(3D-4)^2} \,. \tag{29}$$

We note here that this just implies that the dimensionful quantities flow according to their canonical scaling, and the physically relevant quantity controlling the strength of interactions flows to zero in the infrared [8],

$$\tilde{g}_i^2/\tilde{m}_i^2 \to 0 \,. \tag{30}$$

Further, for a range of initial conditions, only one of the OPs acquires a finite expectation value. In the following discussion we fix the OP with the broken symmetry to be that corresponding to $N_1$, i.e., $\phi$, so that our statement can be rephrased as

$$\lim_{k \to 0} \kappa_{\phi,k} \neq 0 \quad \text{and} \quad \lim_{k \to 0} \kappa_{\chi,k} = 0 \,. \tag{31}$$

This OP can thus be described as flowing to a Nambu-Goldstone infrared fixed point (of that symmetry) where the dimensionless expectation value satisfies

$$\lim_{k \to 0} \kappa_{\phi,k} = \infty \,, \tag{32}$$

while the other OP decouples and flows to a semimetallic-like fixed point (of the other symmetry), i.e, one where Eq. (30) holds.

Finally in the vicinity of the line of exact $O(N)$ symmetry, both OPs acquire a finite expectation value. In particular, this means that crossing from an $O(3)$-broken phase into a $\mathbb{Z}_2$-broken phase takes place as a sequence of transitions in which both symmetries are broken after the first *continuous* transition. This is consistent with the observation in Ref. [9] that the single particle gap at the Dirac points, $\Delta_{\text{sp}} := 2(g_1^2 \rho_\phi + g_2^2 \rho_\chi)$ does not close during the process and changes smoothly across such a transition.

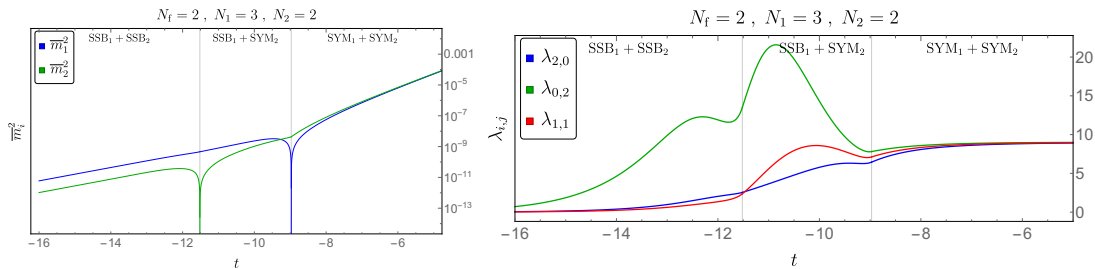

FIG. 2. **Flow trajectories in the coexistence phase.** Upper panel: Evolution of dimensionful transversal boson masses $\overline{m}_i^2$ as a function of RG time $t$ for $N_1 = 3, N_2 = 2$. In the SSB regimes, we plot $\overline{m}_1^2 := \Delta_\phi$ and $\overline{m}_2^2 := \Delta_\chi$. Flows are initialized at $t = 0$ in the symmetric regime close to the IFP. At around $t \approx -9$, the boson mass term $m_1^2 \to 0$ and we switch to the symmetry-broken parametrization where $\kappa_\phi \neq 0$. Then at $t \approx -11.5$ also $m_2^2 \to 0$ and we switch to the parametrization of the boson potential where $\kappa_\phi$ and $\kappa_\chi \neq 0$. Lower panel: RG evolution for the dimensionless quartic couplings exhibiting NG plateaus in $\lambda_{2,0}$ and $\lambda_{1,1}$.

The phase associated to the coexistence region corresponds to an infrared behavior of the system that differs from a situation where all couplings flow together to an infrared fixed point of broken $O(N_1 + N_2)$ symmetry. Moreover, the two OPs do not decouple inside the coexistence fan (as they do in the corresponding regions with only one of the symmetries broken) and this leads to an *approximate $O(N_1 + N_2)$* symmetry. Correspondingly, a snapshot of the joint (normalized) probability density $P(\phi_a, \chi_b)$ for the OPs across the coexistence region does not produce an exact circular pattern but an ellipse. Within our analysis, this can be confirmed from the behavior of the quantity

$$f(\phi_a, \chi_b) := \lim_{k \to 0} V_k(\phi, \chi) \approx f(\phi_a^2 + \chi_b^2) \tag{33}$$

where $V_k$ is the running bosonic potential, and whose level sets coincide with those of $P(\phi_a, \chi_b)$. A conour plot of $f$ is shown in Fig. 3.

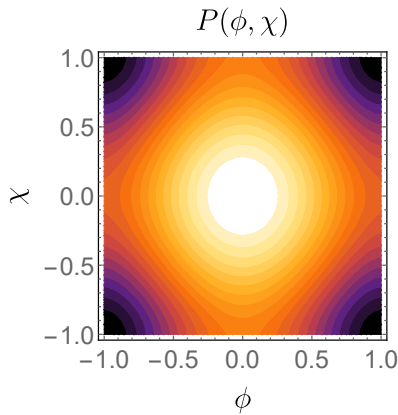

FIG. 3. **Contour plot of the infrared bosonic potential** for $N_1 = 3, N_2 = 1$ for a point inside the coexistence region. The vicinity of the maximum at $\phi, \chi = 0$ has rotational symmetry between the order parameters and gives way to an elliptical shape away from it.

### A. Numerical instability

The presence of Goldstone modes in phases with broken $O(N_i)$-symmetry ($N_i > 1$) leads to the limit $\lim_{k \to 0} \kappa_{i,k}$ being ill-defined numerically close to the line of exact enlarged symmetry. Close to this line, the Goldstone modes force the expectation values of the OPs to switch constantly between zero and nonzero values along the flow in a manner that depends on the finite value of $k$ at which the flow is stopped. That this is indeed the cause of the instability can be confirmed by "turning off" the Goldstone modes, i.e., by considering a model of two coupled Ising

order parameters. In this case the expectation values of both order parameters converge to some nonzero values independent of the scale at which the flow is stopped, meeting continuously at the line of exact $O(2)$ symmetry, as seen in Fig. 4. The residual nonmonotonic behavior as a function of couplings is nonuniversal and depends on the particular path chosen in parameter space.

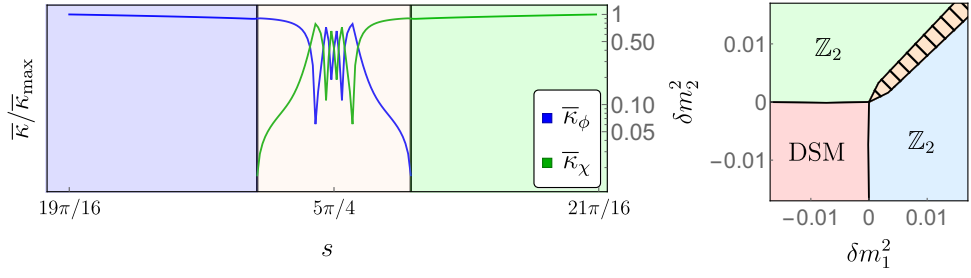

FIG. 4. Left: evolution of the expectation values for $N_1 = N_2 = 1$ along a path in coupling space parametrized by $s$ such that $s = 5\pi/4$ is the line of exact $O(2)$ symmetry. Right: Phase diagram for $N_1 = N_2 = 1$ near the IFP as a function of the tuning parameters $\delta m_i^2$.

## III. DERIVATION OF THE EFFECTIVE QUANTUM SPIN MODEL

The Hamiltonian of the original fermion model from Ref. [9] reads

$$H = H_t + H_U + H_I, \tag{34}$$

with

$$H_t = \sum_{\langle i,j \rangle, s} \left( t + \xi_{ij} \sigma^z_{\langle ij \rangle} \right) c^\dagger_{is} c_{js}, \tag{35}$$

$$H_U = U \sum_i \left( n_{i\uparrow} - \frac{1}{2} \right) \left( n_{i\downarrow} - \frac{1}{2} \right), \tag{36}$$

$$H_I = -J_I \sum_{\langle ij, kl \rangle} \sigma^z_{\langle ij \rangle} \sigma^z_{\langle kl \rangle} - h \sum_{\langle i,j \rangle} \sigma^x_{\langle ij \rangle}, \tag{37}$$

where $s = \uparrow, \downarrow$ is a spin index, and $n_{is} = c^\dagger_{is} c_{is}$. This model is considered at half-filling, and in the large-$U$ limit, we can thus derive an effective quantum spin model within the subspace of singly-occupied sites. To this end, we follow the standard derivation of the Heisenberg model from the Hubbard model [10] by first separating the Hilbert space into subspaces of $m = 0, 1, 2, \ldots$ double occupancies. These subspaces are only connected via the hopping operators $H_t$, which we can split depending on the effect on $m$ into $H_t = T_{-1} + T_0 + T_1$, where $T_j$ changes $m$ to $m + j$. Analogously to the simple Hubbard model, one arrives at the result

$$H_{\text{eff}} = T_0 + H_I - T_{-1} \frac{1}{U + H_I} T_1 + \mathcal{O}\left(\frac{t^3}{U^2}\right) \tag{38}$$

$$= T_0 + H_I - \frac{1}{U} T_{-1} T_1 + \mathcal{O}\left(\frac{t^3}{U^2}, \frac{t^2 J_I}{U^2}\right), \tag{39}$$

which leads to a Heisenberg model $H$ from the main text, with effective couplings that contain Ising spin operators,

$$J^{\text{eff}}_{ij} = \frac{4 \left( t + \xi_{ij} \sigma^z_{\langle ij \rangle} \right)^2}{U} \tag{40}$$

$$= \frac{4 \left( t^2 + \xi_{ij}^2 \right)}{U} + \frac{8 t \xi_{ij}}{U} \sigma^z_{\langle ij \rangle} \tag{41}$$

$$= J_{ij} + \chi_{ij} \sigma^z_{\langle ij \rangle} \tag{42}$$

These relations can thus be used to express the parameters of $H$ in terms of those of the original fermionic model.

# IV. QUANTUM MONTE CARLO SCHEME

In this section, we outline the quantum Monte Carlo (QMC) approach that we used to simulate the effective quantum spin model with the Hamiltonian $H$ from the main text. For this purpose, we used the following hybrid stochastic series expansion (SSE) approach, based on the algorithms introduced in Refs. [11, 12], and working in the $S^z$-$\sigma^z$ direct product basis. The Hamiltonian $H$ can be split into bond operators,

$$-H = \sum_b H^{\rm d}_{\chi,b} + H^{\rm o}_{\chi,b} + H^{\rm d}_{x,b} + H^{\rm o}_{x,b} + H_{z,b} + \sum_{\langle a,b \rangle} H_{{\rm I},a,b} + {\rm const}, \quad (43)$$

with

$$H^{\rm d}_{\chi,b} = (J_b + \chi_b \sigma^z_b)\left(\frac{1}{4} - S^z_{i_b} S^z_{j_b}\right), \quad (44)$$

$$H^{\rm o}_{\chi,b} = (J_b + \chi_b \sigma^z_b)\frac{1}{2}\left(S^+_{i_b} S^-_{j_b} + h.c.\right), \quad (45)$$

$$H^{\rm d}_{x,b} = h, \quad (46)$$

$$H^{\rm o}_{x,b} = h\sigma^x_b, \quad (47)$$

$$H_{z,b} = \epsilon - \frac{\chi_b}{4}\sigma^z_b, \quad (48)$$

$$H_{{\rm I},a,b} = J_I(1 + \sigma^z_a \sigma^z_b), \quad (49)$$

where $a$ and $b$ label nearest-neighbor bonds on the honeycomb lattice. Here, we made use of the bipartiteness of the honeycomb lattice to rotate the sign of the transverse exchange terms $-S^+_i S^-_j \to S^+_i S^-_j$, and choose $\epsilon > \frac{|\chi_b|}{4}$. This decoupling guarantees that the weight of any SSE configuration is always positive, and we thus avoid the sign problem for $|\chi_b| < J_b$. The standard SSE diagonal update [11] provides a way to insert and remove the diagonal operators $H^{\rm d}_{\chi,b}$, $H^{\rm d}_{x,b}$, $H_{z,b}$, and $H_{{\rm I},a,b}$ into the SSE operator sequence.

The above decomposition decouples the Ising and Heisenberg model contributions to the weights of the $H^{\rm d/o}_{\chi,b}$ operators: If the Heisenberg spins are antiparallel, the matrix elements are $(J + \chi_b \sigma^z_b)/2$, whereas for parallel Heisenberg spins, they vanish and the operator does not enter in the operator sequence. This decoupling thus allows us to apply the standard $S = 1/2$ SSE operator loop update [11] in order to efficiently and globally update the Heisenberg sector of the SSE configuration.

It thus remains to construct a global update scheme for the transverse-field Ising model sector of the system. Standard SSE cluster algorithms exist for the transverse-field Ising model [12], which identify clusters of aligned spins in the SSE configuration. However, these are not directly applicable here, because the pure spin inversion symmetry of the uncoupled Ising model sector is explicitly broken for $\chi \neq 0$. One can however still construct clusters by this scheme, and then introduce an a-posteriori acceptance probability, given by the weight ratio of the symmetry breaking operators on the cluster to be flipped,

$$P_{\rm accept} = \min\left\{1, \prod_{p \in {\rm cluster}} \frac{w_p(-\sigma)}{w_p(\sigma)}\right\}. \quad (50)$$

Here, $w_p$ denote the operator weights of the Ising model operators in the constructed cluster, and $\sigma$ the sign of the Ising spins on the cluster (which will be flipped if it is accepted). Typically (e.g., for Ising spins in a uniform longitudinal magnetic field), such an update is expected to perform poorly. In our case, however, the staggered nature of the $\chi$-field can lead to cancellations in the above product. This leads to acceptable performance in the high-$h$ phase of the model, as well as near the quantum critical point. Deeper inside the Ising-ordered phase, $P_{\rm accept}$ decreases and only small clusters are flipped. Therefore, we employ quantum parallel tempering [13–15] in $h$ to ensure the ergodicity of the QMC sampling inside the Ising-ordered phase. Finally, in order to probe for ground-state correlations in the QMC simulations near the quantum critical points, we scaled the SSE simulation temperature inversely proportional to the linear system size, $T = 1/(2L)\, 4(t^2 + \xi^2)/U$, in accord with a dynamical critical exponent $z = 1$ at both quantum phase transitions for $h = h_{c,{\rm I}}$ and $h = h_{c,{\rm H}}$, respectively.

---



\end{document}